\documentclass[prb,preprint,preprintnumbers,amsmath,amssymb,unsortedaddress]{revtex4-1}

\usepackage{graphicx}% Include figure files
\usepackage{dcolumn}%Align table columns on decimal point
\usepackage{bm}% bold math
\usepackage{color}
\usepackage{float}
\makeatletter
\newcommand{\Rmnum}[1]{\expandafter\@slowromancap\romannumeral #1@}
\newcommand{\jgr}{J. Geophys. Res. }
\newcommand{\aap}{Astron. Astrophys. }
\newcommand{\ssr}{Space Sci. Rev. }
\newcommand{\aapr}{Astron. Astrophys. Rev. }

\newcommand{\solphys}{Sol. Phys. }
\newcommand{\grl}{Geophys. Res. Lett. }
\makeatother

\begin{document}

\title{The effect of electron holes on cyclotron maser emission driven by horseshoe distributions}

\author{G. Q. Zhao}
\affiliation{Institute of Space Physics, Luoyang Normal University, Luoyang, China}
\author{Y. H. Chu}
\affiliation{Institute of Space Science, National Central University, Chungli, Taiwan}
\author{H. Q. Feng}
\affiliation{Institute of Space Physics, Luoyang Normal University, Luoyang, China}
\author{D.~J.~Wu}
\affiliation{Purple Mountain Observatory, CAS, Nanjing, China}

\renewcommand{\abstractname}{}
\begin{abstract}
~\\
This Brief Communication presents a quantitative investigation for the effect of electron holes on electron-cyclotron maser (ECM) driven by horseshoe distributions. The investigation is based on an integrated distribution function for the horseshoe distributions with electron holes. Results show that the presence of electron holes can significantly enhance the ECM growth rate by 2$\sim$3 times in a very narrow waveband. The present study suggests that these electron holes probably are responsible for some fine structures of radiations, such as narrowband events in auroral kilometric radiation and solar microwave spikes.
\end{abstract}

\maketitle

%\newpage
%\section{INTRODUCTION}
It is well acknowledged that a horseshoe distribution will be formed when a beam of electrons moves along the magnetic field with enhanced field strength. The distribution can efficiently drive electron-cyclotron maser (ECM) in a strongly magnetized plasma, which has been extensively discussed in the context of Earth's auroral kilometric radiation, and radio emissions from the Sun and astrophysical shocks. \citep{wuc85p15,tre06p29,bin13p95} On the other hand, electron holes are believed to be a common phenomenon in various plasma environments, particularly in auroral zone. \citep{mat94p15,erg98p25,bal98p29,cat99p25,and09p04,mal13p91,moz16p01} It has been proposed that the presence of electron holes may distort the horseshoe distribution, and therefore modulate ECM. \citep{pot01p65,pot05p04,tre11p85,tre12p19} However, to the best of our knowledge, some quantitative study on ECM driven by the horseshoe distribution with electron holes is not well documented. The study requires a specific form of the distribution function describing electron holes in two-dimensional momentum space. \citep{tre11p85,tre12p19}

Based on a specific distribution function for the horseshoe distribution with an electron hole, this Brief Communication investigates the effect of the electron hole on ECM driven by horseshoe distributions. Results show that the presence of electron holes can significantly affects the ECM, which first enhances the ECM growth rate. Moreover, the frequency range with enhancement of growth rate is much narrower than that with conventional growth without electron holes. These results provide evidence to support the idea that electron holes contribute to the fine structure of the auroral kilometric radiation. \citep{pot01p65,pot05p04,tre11p85,tre12p19}
%, from the kinetic view,

Let us first model the distribution function for the horseshoe distribution. We consider that the non-thermal electron distribution has a power-law energy spectrum with lower energy cutoff. \citep{ree69p13,hon71p63,gan01p58,lin11p21} We also introduce the hyperbolic tangent function with parameters of steepness index $\delta$ and cutoff energy $E_c$, which is convenient to describe the lower energy cutoff. \citep{zha13p75,zha16p58,zha16p09} The steepness index determines the steepness when the particle population rises with respect to energy, while cutoff energy denotes the energy with a population inversion followed by the power-law spectrum. Further, the angular distribution of the non-thermal electrons is considered to be characterized by a Gaussian distribution. \citep{lee00p57}
Consequently, the distribution function is represented as
\begin{eqnarray}
f(u,\mu)=A\tanh {{(u/u_{c})}^{2\delta}}u^{-2\alpha-1}\exp{[-\frac{(\mu-1)^2}{\Delta\mu^2}]},
\end{eqnarray}
where $u$ is the momentum per unit mass, $\mu$ the pitch-angle cosine of electrons, $A$ is the normalized factor, $u_c$ is the corresponding cutoff momentum determined by $E_c$, $\alpha$ is the power index, and $\Delta\mu$ is the half-width in pitch angles.
Note that Eq. (1) can describe the horseshoe distribution for $\delta > \alpha$ as well as a large $\Delta\mu$ ($\gg 0.1$), \citep{zha16p58} which will be shown in Fig. 1. %the momentum distribution in Eq.(1) has a population inversion around $u=u_c$ when $\delta > \alpha$, and

It is very difficult to obtain an exact electron distribution for the holes appearing in a horseshoe distribution. For simplicity holes in the present paper are considered to be small regions of lacking particles in the momentum space. \citep{tre11p85,tre12p19} Based on this concept we can represent the horseshoe distribution with a hole by making full use of Eq. (1) with the form
\begin{eqnarray}
F(u,\mu)=f(u,\mu)-rf^*(u,\mu),
\end{eqnarray}
and
\begin{eqnarray}
f^*(u,\mu)=A\tanh {{(u/u_{c}^*)}^{2\delta^*}}u^{-2\alpha^*-1}\exp{[-\frac{(\mu-1)^2}{\Delta\mu^{*2}}]}. \nonumber
\end{eqnarray}
The first term on the right hand side of Eq. (2) corresponds to the horseshoe distribution, while the other term is used to determine the hole. Here $r$ is a constant to make sure that Eq. (2) is nonnegative in the whole momentum space, and $r=0$ implies no electron hole. The parameters marked by $*$ have the same meanings as those without $*$, but may have different parameter values.

Fig. 1 displays the distribution function described by Eq. (2) in the two-dimensional momentum space, where $u_z$ and $u_\bot$ are the components of the vector $\bf{u}$ parallel and perpendicular to the ambient magnetic field, and the momentum has been normalized by the speed of light $c$. One may first find a horseshoe distribution in Fig. 1. To obtain the horseshoe distribution, we have set the parameters $\alpha=2$, $\delta=4$ and $\Delta\mu=1$ that are adopted from the literature. \citep{ree69p13,men78p75,lin82p49} We have also set $E_c=5 $ keV which is a typical value for auroral energetic electrons. Another feature shown in Fig. 1 is the presence of a small region of lacking electrons in the horseshoe distribution, namely ``electron hole" defined in the present paper. This electron hole is obtained by setting the parameters $\alpha^*=8$, $\delta^*=16$, $\Delta\mu^*=0.01$, $E_c^*=4.5$ keV, and $r=0.7$. Here we emphasize that the choice of these parameter values is subjective and other choices that may lead to different shapes, sizes, and/or positions of holes are possible. It is worthy to note that Eq. (2) can conveniently describe the horseshoe distribution with an electron hole. This allows us to investigates the effect of such a hole on the ECM through calculating ECM growth rates numerically.

The general formula of ECM growth rates is well known and given by \citep[e.g.,][]{mel86,che02p16}
\begin{eqnarray}
\gamma_{\sigma} & = &
 \frac{\pi}{2}\frac{n_{b}}{n_{0}}\frac{\omega_{pe}^{2}}{\omega}
\frac{1}{(1+T_{\sigma}^{2})R_{{\sigma}}}\int{d^{3}\bf
u}\gamma(1-\mu^{2})\delta\left(\gamma-\frac{s\Omega_{e}}
{\omega}-\frac{N_{{\sigma}}u\mu}{c}\cos\theta\right) \nonumber \\
 & & \times\left\{\frac{\omega}{\Omega_{e}}\left[\gamma\text{$K_{{\sigma}}$}\sin\theta+T_{{\sigma}}\left(\gamma\cos\theta-\frac{N_{{\sigma}}u\mu}{c}\right)\right]
\frac{J_{s}(b_{{\sigma}})}{b_{{\sigma}}}+J_s^{'}(b_{{\sigma}})\right\}^{2} \nonumber \\
 & & \times\left[u\frac{\partial}{\partial\text{$u$}}+\left(\frac{N_{{\sigma}}u\cos\theta}{{\gamma}c}-\mu\right)\frac{\partial}{\partial\mu}\right]F(u,\mu),
\end{eqnarray}
with
\begin{eqnarray}
b_{{\sigma}}& = & N_{{\sigma}}({\omega}/\Omega_{e})(u/c)\sqrt{1-\mu^{2}}\sin\theta,\nonumber \\
R_{{\sigma}} & = &
1-\frac{\omega_{pe}^{2}\Omega_{e}\tau_{{\sigma}}}{2{\omega}({\omega}+\tau_{{\sigma}}\Omega_{e})^{2}}\times
\left(1-{\sigma}\frac{s_{{\sigma}}}{\sqrt{s_{\sigma}^{2}+\cos^{2}\theta}}\frac{\omega^{2}+\omega_{pe}^{2}}{\omega^{2}-\omega_{pe}^{2}}\right),\nonumber \\
K_{{\sigma}} & = & \frac{\omega_{pe}^{2}\Omega_{e}\sin^{2}\theta}{(\omega^{2}-\omega_{pe}^{2})({\omega}+\tau_{{\sigma}}\Omega_{e})},\nonumber \\
T_{{\sigma}} & = & -\frac{\cos\theta}{\tau_{{\sigma}}}, \nonumber
\\
\tau_{{\sigma}}& = &-s_{{\sigma}}+{\sigma}\sqrt{s_{{\sigma}}^{2}+\cos^{2}\theta}, ~~~~
s_{{\sigma}}=\frac{{\omega}\Omega_{e}\sin^{2}\theta}{2({\omega}^{2}-\omega_{pe}^{2})}.
\end{eqnarray}
Here $n_0$ and $n_{b}$ denote electron number densities of the ambient plasma and non-thermal electrons; $\omega_{pe}$ is the plasma frequency and $\Omega_e$ is the electron gyrofrequency; $\gamma=\sqrt{1+u^2/c^2}$ is the relativistic factor; ${J_s}(b_{{\sigma}})$ and $J_s^{'}(b_{{\sigma}})$ are the Bessel function of the order of $s$ and the derivative, respectively; $\sigma=+$ is for the O-mode and $\sigma=-$ for the X-mode; $\omega$ is the emitted wave frequency, and $\theta$ is the propagation angle of the emitted waves with respect to the ambient magnetic field; ${N_ \sigma }$ is the refractive index determined approximately by cold-plasma theory.

On the basis of Eqs. (2) and (3), one can calculate the growth rates of radiations in both O-mode and X-mode once a ambient plasma parameter is known. The plasma parameter is the frequency ratio of the plasma frequency to electron gyrofrequency, i.e., $\omega_{pe}/\Omega_e$. In the present study $\omega_{pe}/\Omega_e$ = 0.01 shall be considered since it is a typical value in the auroral zone. The growth rate depends on two variables ($\omega$, $\theta$). One can obtain the peak growth rate referring to the highest magnitude as a function of one variable with the other fixed, and further obtain the maximum growth rate with the highest values in both variables. Under the condition $\omega_{pe}/\Omega_e$ = 0.01, fundamental X-mode is the fastest growth mode according to our calculations via comparing the maximum growth rates of radiations with each other for various wave modes. Hence the following discussions pay attention only to the fundamental X-mode.

Fig. 2 plots the peak growth rate of the fundamental X-mode with respect to the propagation angle. The growth rate has been normalized by $\Omega_e n_b/n_0$. Panels (a)$-$(c) correspond to three cases of electron holes determined by $E_c^*=4, 4.5$, and 5 keV, respectively. Different values of $E_c^*$ represent different positions of holes in the momentum space. In each panel the solid line is for the horseshoe distribution with an electron hole, while the dashed line (marked in red for comparison) is for that without electron hole. From each panel, one may first find that the maximum growth rate with the effect of electron hole is much larger than that without the electron hole. The range of the propagation angle with wave growth also becomes wide when the effect of electron hole is taken into account, a larger $E_c^*$ leads to a wider angle range in Fig. 2. It should be noted that the propagation angles with maximum growth are different for different situations.
The maximum growth rate with the effect of electron hole takes place at about $82.4^{\circ}$ in the case of $E_c^*=4.5$ keV (panel (b)), while it takes place near $84.1^{\circ}$ if without the electron hole. For a given propagation angle, the peak growth rate is often larger than that in the absence of electron hole when the wave is emitted. It is clear that the presence of electron holes enhances the ECM growth in a large propagation angle range.

To reveal the frequency property of radiations with the effect of electron holes, Fig. 3 presents the ECM growth rate versus the emitted wave frequency for a given propagation angle $\theta=83.3^{\circ}$. Similar to Fig. 2, Panels (a)$-$(c) in Fig. 3 correspond to three cases of $E_c^*=4, 4.5$, and 5 keV, respectively, and in each panel the solid line is for the horseshoe distribution with an electron hole while the red dashed line is for that without electron hole. A prominent result is that the enhancement of the ECM growth rate does not happen in the whole frequency range with wave growth. On the contrary, it happens only in a very narrow frequency range relative to the range of wave growth driven by the global horseshoe distribution. This result is reasonable because the size of the hole is much smaller than that of the horseshoe distribution in momentum space. In addition, the enhancement takes place at high frequency part of the frequency band with wave growth, and the frequency range with growth rate enhancement becomes narrower for a lower $E_c^*$ (panel (a)).

The present study is motivated mainly by the phenomenon of Earth's auroral kilometric radiation which was first investigated comprehensively by Gurnett. \citep{gur74p27} The radiation is characterized by a very high brightness temperature ($>10^{15}$ K) implying a coherent emission process. It can occur in the frequency range of about 50$-$500 kHz, and appears as continuum emission sometimes lasting for several hours. \citep{gur74p27,lou06p55}
It is now generally believed that this continuum radiation is produced by ECM instability driven by horseshoe electron distributions. The distributions result from potential-accelerated electrons moving along an enhanced magnetic field in auroral zone. In particular, based on high time and frequency resolution measurements, Pottelette et al. \citep{pot01p65} found that the auroral kilometric radiation consists of a large number of elementary radio events, which superpose on the continuum radiation background or even may compose all the radiation. Each elementary event has an extremely narrow bandwidth, i.e., $\triangle f \approx 0.1$ kHz for the radiation at frequency about 400 kHz. Such narrow bandwidth is smaller by roughly one order of magnitude and, as argued by Pottelette et al., \citep{pot01p65} can hardly be explained based on ECM driven by the global horseshoe distribution. To find the radiation origin characterized by such narrow bandwidth, the authors qualitatively proposed that it is attributed to electron hole embedded into an unstable horseshoe electron distribution. Each electron hole was expected to contribute to stimulate radio emission by the interaction of electron hole with horseshoe electron distribution, i.e., cutting out a certain region in electron phase space, which decreases the density of the region and introduces a positive perpendicular gradient to drive ECM. It is clear that the present quantitative study supports this idea, since the results show the presence of electron holes considerably enhance the ECM growth rate in a very narrow waveband.
%The elementary radio events were interpreted as the signature of the interaction of electron holes with the horseshoe electron distribution. \citep{pot01p65,pot05p04,tre11p85,tre12p19} %with short lifetimes ($< 1 s$)

In addition, other narrowband radio events from astrophysical objects may also be relevant. Solar microwave spikes, for instance, have gained much attention because they possibly provide direct signature of the elementary flare processes such as electron acceleration and energy release. \citep{ben86p99}
These spikes usually appear in clusters of individual events which are randomly distributed in radio dynamic spectra. They are narrowband bursts with a very high brightness temperature ($\sim 10^{15}$ K) and the smallest bandwidth relative to the center frequency is $0.17\%$ only. \citep{mes00p87} (Note that the bandwidth can be broadened if the plasma in spike source is inhomogeneous.) A lot of investigators believe that they are produced via ECM mechanism, in which a horseshoe or loss cone electron distribution is invoked. \citep{mel82p44,fle03p71,cli11p45} However, the present study perhaps can give a new insight into such radio events. Solar flares probably serve for the generation of electron holes since they are highly complex processes with magnetic field reconnection, particle acceleration, plasmoid ejection, double layer formation, and various kinetic activities. The narrowband spike bursts may be expected if many small-scale electron holes are produced in a solar flare. We believe that the effect of electron holes should be helpful for the generation of such narrowband bursts. Further discussion on this issue is also important but is beyond the scope of the present Brief Communication.

Before concluding, we remark several points. First, the enhancement of ECM growth rates in the present study is due to the density depression of small-size electron holes embedded into the horseshoe electron distribution. The density depression contributes to a positive perpendicular gradient on the distribution function and therefore drives ECM. It is clear that the steepness of this positive gradient determines the degree of the enhancement of ECM growth rates. The electron hole shown in Fig. 1 has a size of $\Delta u_z/u_c^* \simeq 0.24$ and $\Delta u_{\bot}/u_c^* \simeq 0.34$, where $\Delta u_z$ ($\Delta u_\bot)$ is the momentum difference between the boundaries of the hole in parallel (perpendicular) direction with respect to the ambient magnetic field, and $u_c^*$ represents the momentum of the hole determined by $E_c^*$. The density depression in such a size is about $0.1\%$ relative to the electron density of the horseshoe distribution. A larger density depression will result in a larger enhancement of ECM growth rates, and vice versa based on our calculations.
Second, the perpendicular gradient can be expressed as the sum of two terms, i.e., $\frac{\partial F}{\partial u_{\bot}} = \frac{\cos \phi}{u} \frac{\partial F}{\partial \phi}+\sin \phi \frac{\partial F}{\partial u} $, where $\phi$ is the electron pitch angle. \cite{pot01p65} Particularly, our calculations also show that the emissions due to the electron holes are mainly determined by the first term. This should be reasonable since the holes in the present study are located at very small values of $\phi$ in the momentum space, and thus the term $\sin \phi \frac{\partial F}{\partial u}$ can be neglected. \cite{pot14p77}
Third, our study does not consider the movement of electron holes in configuration space. The movement will lead to a frequency drift of the emitted spectrum due to, at least, the change of the ambient magnetic field. The drift rate is determined mainly by both the movement velocity of electron holes and the distribution of the ambient magnetic field. The velocity used in this paper is around $3.8 \times 10^4$ km/s, which may result in a very large drift rate exceeding 4000 kHz/s based on the assumption of a dipole magnetic field geometry. \cite{pot07p35} A recent comprehensive study has revealed that there exist intense emissions of auroral kilometric radiation associated simultaneously with the presence of electron holes exhibiting predominantly very large drift rates. \cite{pot14p77} (Unfortunately authors did not give the drift rates of these events because they are too large to be accurately determined due to the experimental constraints.) Moreover, the study \cite{pot14p77} has also found that the associated electron distribution presents a pronounced beam-like shape. According to a recent study by Zhao et al., \cite{zha16p09} the ECM with the effect of electron beams is dominated possibly by the O-mode as well as by the X-mode, which should be investigated further in a future study.

In summary, this Brief Communication models a horseshoe distribution with electron hole for the first time and quantitatively calculates its ECM growth rates. We propose that the presence of electron holes should add the total emission from the horseshoe distribution and contribute to narrowband fine structure, which coincides with the appearance of narrowband events in auroral kilometric radiation and even solar microwave spikes. This Brief Communication is preliminary and more in-depth study is desirable.

~\\
%\acknowledgments
G. Q. Zhao acknowledges the National Central University in Taiwan for hospitality during his recent visit.
Research by G. Q. Zhao and H. Q. Feng was sponsored by NSFC under grant Nos. 41504131, 41274180, and 41231068, and was also supported partly by the Science and Technology Project of Henan Province (Grant Nos. 17A170002, 142102210109 and 13IRTSTHN020). Research by D. J. Wu was supported partly by NSFC under grant Nos. 41531071, 11373070. The authors are grateful to the anonymous referee for valuable comments and suggestions.

\newpage
%\bibliography{my_ref}

\newpage
\begin{figure}
\includegraphics[scale=1.0]{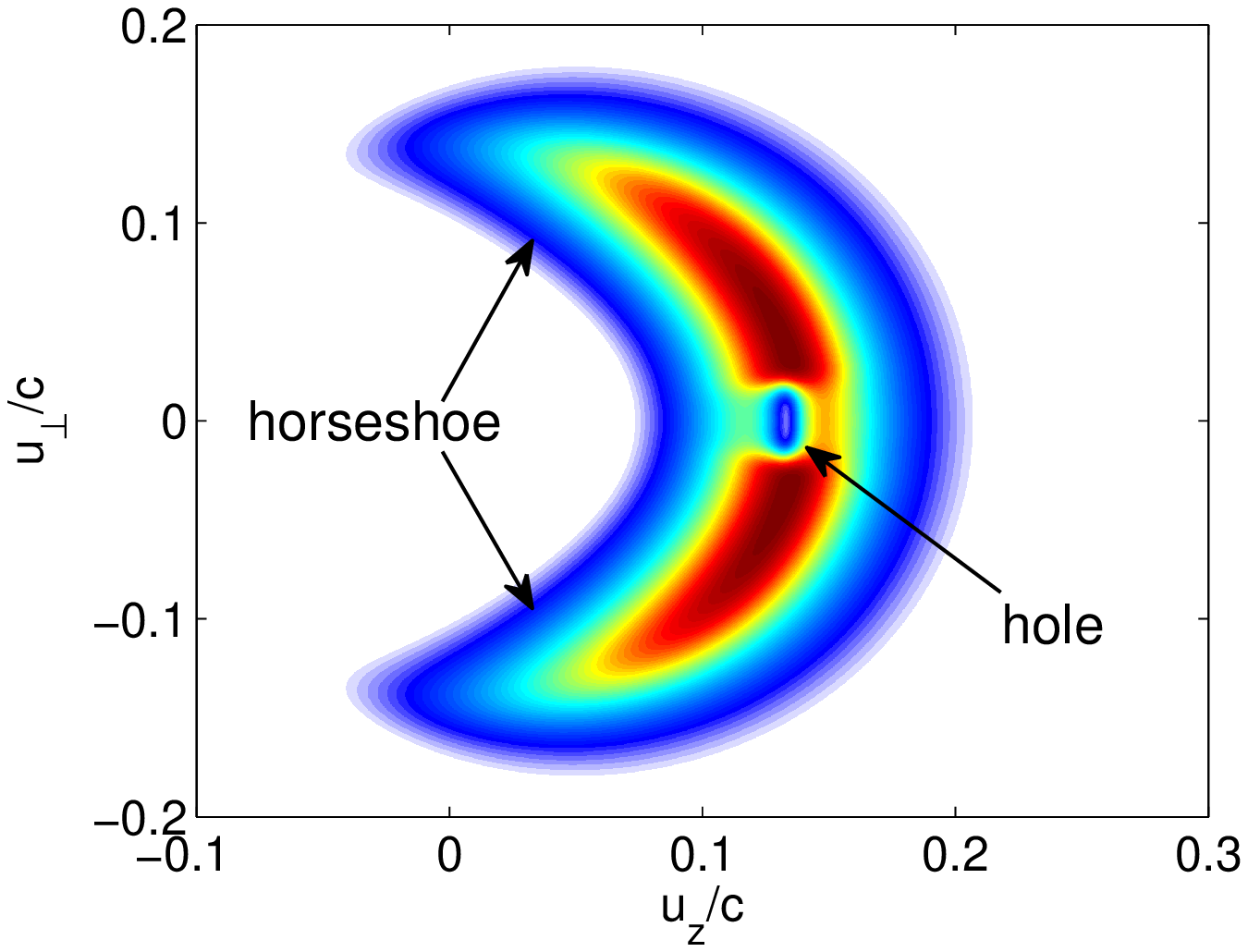} \caption{Colormap plot of the distribution function described by Eq. (2). It is clear that Eq. (2) describes a horseshoe distribution with an electron hole.\label{fig1}}
\end{figure}

\begin{figure}
\includegraphics[scale=1.2]{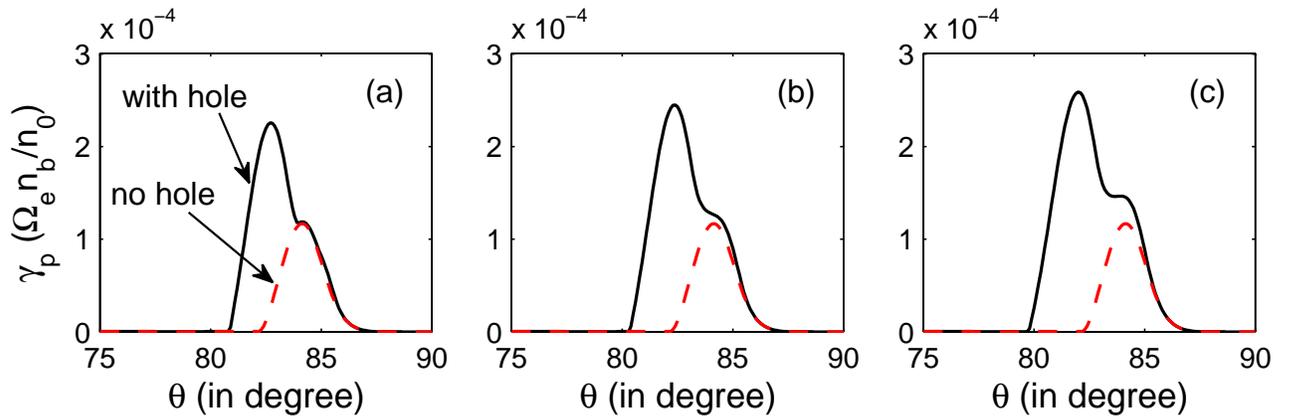} \caption{Peak growth rate of the fundamental X-mode versus the propagation angle $\theta$. Three panels correspond to three diffrent positions of electron holes in the momentum space determined by $E_c^*=4, 4.5$, and 5 keV, respectively. In each panel the solid line is for the horseshoe distribution with an electron hole, while the red dashed line is for that without electron hole. \label{fig2}}
\end{figure}

\begin{figure}
\includegraphics[scale=1.2]{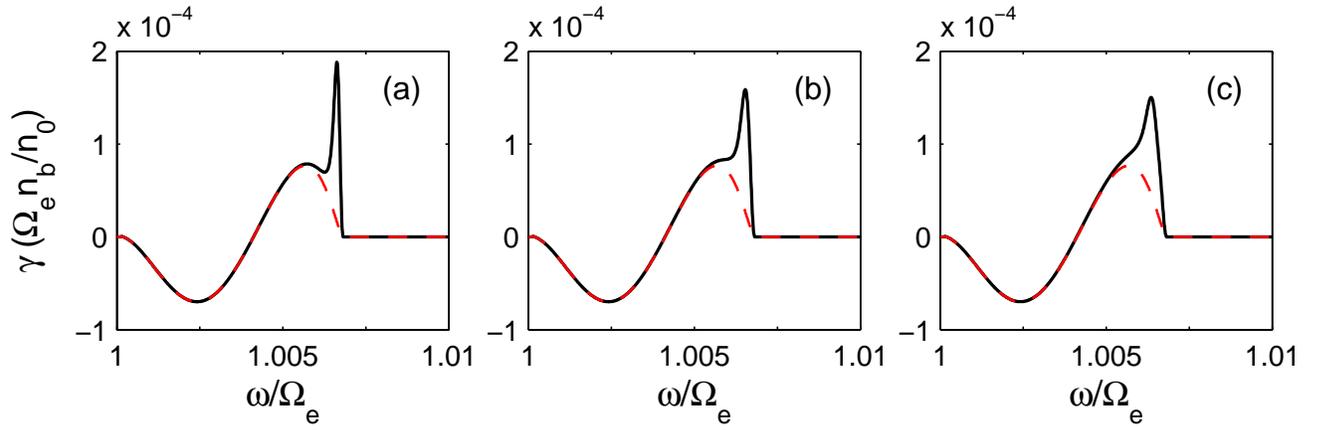} \caption{Growth rate of the fundamental X-mode versus the normalized frequency. The implication of panels and lines are the same as in Fig. 2. One can find that the narrow frequency band with larger growth rate appears in each panel.  \label{fig3}}
\end{figure}

\end{document}